\documentclass[12pt]{iopart}
\usepackage{iopams}  
\begin{document}

\title[Hydrodynamic Nambu Brackets derived by Geometric Constraints
]{Hydrodynamic Nambu Brackets derived by Geometric Constraints
}

\author{Richard Blender}
\address{Meteorological Institute, University of Hamburg, 
Hamburg, Germany}
\ead{richard.blender@uni-hamburg.de}
\author{Gualtiero Badin}
\address{Institute of Oceanography, University of Hamburg, Hamburg, Germany}
\ead{gualtiero.badin@uni-hamburg.de}	


\begin{abstract}
A geometric approach to derive the Nambu brackets for ideal 
two-dimensional (2D) hydrodynamics is suggested. The derivation is based on two-forms with vanishing integrals in a periodic domain, and with resulting dynamics constrained by an orthogonality condition. As a result, 2D hydrodynamics with vorticity as dynamic variable emerges as a generic model, with conservation laws which can be interpreted as enstrophy and energy functionals. 
Generalized forms like surface quasi-geostrophy and fractional 
Poisson equations for the stream-function are also included as results from the derivation. 
The formalism is extended to a hydrodynamic system coupled to a second degree of freedom, with the Rayleigh-B\'{e}nard convection as an example.
This system is reformulated in terms of constitutive conservation laws with two additive brackets which represent individual processes: a first representing inviscid 2D hydrodynamics, and a second representing the coupling between hydrodynamics and thermodynamics. The results can be used for the formulation of conservative numerical algorithms that can be employed, for example, for the study of fronts and singularities. 

\end{abstract}


\today

\maketitle

\section{Introduction}

In a seminal article, Nambu \cite{Nambu1973} suggested an extension of Hamiltonian dynamics which is based on Liouville's Theorem 
and, differently from classical Hamiltonian mechanics, makes use of several conserved quantities. 
The additional conservation laws (CLs) can be considered as additional Hamiltonians and define manifolds whose intersection determine the trajectory in state space. In analogy to Poisson brackets in Hamiltonian mechanics, the resulting dynamics is determined by Nambu brackets \cite{Takhtajan1994}. 
As a first example, Nambu considered the Euler equations for the rotating solid body \cite{Nambu1973}. The approach has then been applied to a variety of finite-dimensional systems, ranging from the nondissipative Lorenz equations \cite{NevirBlender1994} to the geometry of strange attractors in dissipative, chaotic systems \cite{Roupas2012}. For early examples of the derivation of the Nambu equations for different systems, see \cite{Chatterjee1996,Cohen1975}.

The Nambu approach was extended from finite to infinite dimensional systems, and in particular to hydrodynamics, by N\'{e}vir and Blender \cite{NevirBlender1993}, who were able to determine the Nambu brackets corresponding to ideal hydrodynamics in two and three dimensions, with enstrophy and helicity as CLs which exist as consequence of the particle relabeling symmetry \cite{Salmon2008}.
An important finding was that  a Nambu representation can be useful to construct numerical algorithms for the simulation of geophysical flows  \cite{Salmon2005, Salmon2007}.
The application of this approach in a global shallow water model 
revealed a distinct impact on energy and enstrophy spectra \cite{SommerNevir2009}. 
In the past years, several models of geophysical fluid dynamics  
have been rewritten in terms of Nambu brackets \cite{Bihlo2008,GassmannHerzog2008,NevirSommer2009,
SalazarKurgansky2010,BlenderLucarini2013}. 

%
%

It should be noted that, apart for notable exceptions (e.g. \cite{Sommeretal2011}) the construction of the hydrodynamics Nambu brackets was mainly based on intuition and guessing. The aim of this study is thus to derive the Nambu brackets for hydrodynamics using a geometrical approach, which is based on replacing the Jacobian by a two-form based on the two CLs, as suggested for finite dimensional systems by  \cite{Fecko1992}. While the examples here analysed have already been studied in the past, we report a novel way to systematically derive the hydrodynamics Nambu brackets.
The method is introduced in Section \ref{sec:G3D} for a physical system with three degrees of freedom and two CLs. 
In Section \ref{sec:H2D}, the geometrical approach is used to derive the Nambu bracket for 2D hydrodynamics. 
The Nambu form for 2D hydrodynamics emerges when the dynamic variable is interpreted as vorticity, and the two CLs are the kinetic energy and the enstrophy. 
The resulting dynamics can be extended to the case of generalized Euler equations
with a fractional Poisson operator for the stream-function. 
In Section \ref{sec:coupled}, the approach is extended to a coupled system with
 two degrees of freedom. Rayleigh-B\'{e}nard convection is discussed as an example. 
Finally, in Section \ref{sec:ccl} the method is used to construct a Nambu representation of the coupled model in terms of constitutive conservation laws, i.e. CLs which are conserved only in sub-systems. 

\section{Geometry of Nambu mechanics in 3D}
\label{sec:G3D}

Consider a physical system with three degrees of freedom, $X=(x_1,x_2,x_3)$ and two CLs, $H = H(X)$ and $C = C(X)$. Since $H$ and $C$ are invariant,
\begin{equation}
\frac{d H}{d t} = \nabla_{X} H \cdot \frac{d X}{d t} = 0 ,
\label{eq:21}
\end{equation}		
\begin{equation}
\frac{d C}{d t} = \nabla_{X} C \cdot \frac{d X}{d t} = 0 ,
\label{eq:22}
\end{equation}
where $\nabla_{X}$ indicates the gradient operator in the $X$ space. 
(\ref{eq:21})-(\ref{eq:22}) are satisfied if the vector $dX / dt$  is orthogonal to  $\nabla_X H$  and  $\nabla_X C$. 
An orthogonal vector can be constructed using the three-dimensional vector product
\begin{equation}
\frac{d X}{d t} = \nabla_X C \times \nabla_X H ,
\label{eq:23}
\end{equation}	
which shows that the flow in phase space $dX / dt$ is along the intersection of the manifolds defined 
by constant $H$ and $C$. Equation (\ref{eq:23}) is the canonical Nambu form
\begin{equation}
\frac{d x_n}{d t} = \frac{\partial \left(x_n,C,H \right)}{\partial \left( x_1 , x_2 , x_3 \right) } ,~n=1,2,3,
\label{eq:24}
\end{equation}
Equation (\ref{eq:24}) can be rewritten using the anti-symmetric Levi-Civita symbol $\varepsilon_{nij}$ as
\begin{equation}
\frac{d x_n}{d t} = \varepsilon_{nij} \frac{\partial C}{\partial x_i}\frac{\partial H}{\partial x_j}  , \qquad n,i,j=1,2,3.
\label{eq:25}
\end{equation}
The dynamics of an arbitrary state space function $F(X)$ is thus given by the canonical Nambu bracket
\begin{equation}
\frac{\partial F}{\partial t} = \left\{ F,C,H\right\} .
\label{eq:26}
\end{equation}
Geometrically, the bracket is the volume of the parallelepiped
\begin{equation}
\left\{ F,C,H\right\} = \nabla_X F \cdot \nabla_X C \times \nabla_X H ,
\label{eq:27}
\end{equation}
The conservation of $H$ and $C$ is obtained from (\ref{eq:27}), since the volume vanishes for $F=H$ or $F=C$. 

As stated in the Introduction, the Nambu bracket differ from the usual Poisson bracket for being defined on a phase space that can be odd dimensional and of higher dimensionality, i.e. of dimension $n \ge 3$. If (\ref{eq:27}) is generalized as $\{f_1,...,f_n\}$, the bracket is a multilinear map
\begin{equation}
\left\{ ~,~,~\right\} ~:~[C^{\infty}(X)]^{\otimes n} \rightarrow C^{\infty}(X),
\label{eq:27a}
\end{equation}
$\forall f_i~(i=1,...,n) \in X$, where $X$ is a smooth manifold. (\ref{eq:27a}) satisfies the following properties:
\begin{itemize}
\item {\it Skew-symmetry} 
\begin{equation}
\left\{ f_1,\cdots,f_n\right\} =(-1)^{\varepsilon (\sigma)} \left\{ f_{\sigma_1},\cdots,f_{\sigma_n}\right\},
\label{eq:27b}
\end{equation}
where $\varepsilon  (\sigma)$ is the parity of a permutation $\sigma$.
\item {\it Leibniz Rule} 
\begin{equation}
\left\{ f_1,\cdots,f_n\right\} =f_1 \left\{ f_2,\cdots,f_n\right\}+\left\{ f_1,f_3,\cdots,f_n\right\}f_2.
\label{eq:27c}
\end{equation}
\item {\it Fundamental Jacobi identity} 
\begin{eqnarray}
\{\{f_1,...,f_{n-1},f_n\},f_{n+1},...,f_{2n-1}\}+\\ \nonumber
+\{f_n,\{f_1, ...,f_{n-1},f_{n+1}\},f_{n+2},...,f_{2n-1}\}+ \cdots \\ \nonumber
+ \{f_n,...,f_{2n-2},\{f_1,...,f_{n-1},f_{2n-1}\}\} \\ \nonumber
= \{f_1,...,f_{n-1},\{f_n,...,f_{2n-1}\}\}.
\label{eq:27d}
\end{eqnarray}
\end{itemize}
For more algebraic properties of Nambu brackets see \cite{Takhtajan1994}. Finally, it should be noted that, due to the multiple Hamiltonian structure of Nambu mechanics, Nambu systems have the property of possessing dynamical or hidden symmetries resulting in extra integrals of motion beyond those needed for complete integrability \cite{Chatterjee1996}.  

\section{Hydrodynamics in 2D}
\label{sec:H2D}


\subsection{Geometric derivation of the Nambu bracket}

Consider a 2D continuous system with
the dynamic variable  $\zeta$ and the two 
conserved functionals $H = H[\zeta]$ and
$E = E[\zeta]$. The conservation of $H$ and $E$ yields 
%
\begin{equation}
\frac{d H}{d t} 
= \int H_{\zeta} \frac{\partial \zeta}{\partial t} dA = 0 ,
\label{eq:31}
\end{equation}
\begin{equation}
\frac{d E}{d t} 
= \int E_{\zeta} \frac{\partial \zeta}{\partial t} dA = 0 ,
\label{eq:32}
\end{equation}
where  $H_{\zeta} = \delta H / \delta \zeta$ and $E_{\zeta} = \delta E / \delta \zeta$   are the functional derivatives of $H$ and $E$ with respect to $\zeta$, and $dA=dx_1~dx_2$
is an area element. In analogy to the 3D state space considered
in Section \ref{sec:G3D},
the sums (\ref{eq:21})-(\ref{eq:22}) are replaced by the integrals in the $(x_1,x_2)$ plane. In (\ref{eq:31})-(\ref{eq:32}), the CLs are constraints for the time evolution of $\zeta$ and can be interpreted as an orthogonality condition for $E_{\zeta}$  and  $H_{\zeta}$  with respect to  $\partial \zeta / \partial t$. 

To construct dynamics which satisfies these orthogonality relations, we use differential forms (see, e.g., \cite{Arnold89}). Consider a 2-form $df \wedge dg$ for arbitrary functions  $f$  and $g$, which is exact, so that $df \wedge dg  = d(f \wedge dg)$.
The integral of the 2-form thus vanishes in a periodic domain 
\begin{equation}
\int df \wedge dg =0 .
\label{eq:33}
\end{equation}
Equation (\ref{eq:33}) yields   
\begin{equation}
\begin{array}{ccc}
\int f ~ df \wedge dg =0 & and & \int g ~ df \wedge dg =0 ,
\end{array}
\label{eq:34}
\end{equation}
which shows that  $f$  and  $g$ are orthogonal to the 2-form  $df \wedge dg$. 
With the identifications 
$f = H_{\zeta}$, 
$g = E_{\zeta}$, 
(\ref{eq:31})-(\ref{eq:32}) suggests
\begin{equation}
\frac{\partial \zeta}{\partial t} ~ dA = df \wedge dg = J (f,g) dx_1 \wedge dx_2 ,
\label{eq:35}
\end{equation}
where  $J$  is the Jacobian $J(f,g) = f_{x_1} g_{x_2} - f_{x_2} g_{x_1}$ with the subscripts indicating partial derivatives.
The orthogonality constraint (\ref{eq:35}) yields 
\begin{equation}
\frac{\partial \zeta}{\partial t} = J \left( H_{\zeta},E_{\zeta} \right) 
\label{eq:36}
\end{equation}
which is the Nambu representation of 2D
hydrodynamics \cite{NevirBlender1993}.
Arbitrary functionals $F[\zeta]$  evolve according to
\begin{equation}
\frac{\partial F}{\partial t} = \{ F,E,H \} ,
\label{eq:37}
\end{equation}
with the Nambu bracket 
\begin{equation}
\{F,E,H\} = - \int F_{\zeta} J \left( E_{\zeta}, H_{\zeta} \right) dA .
\label{eq:38}
\end{equation}
%
%
%
The bracket (\ref{eq:38}) is anti-symmetric and cyclic, $\{ F,E,H \} = \{E,H,F \} = \{H,F,E \}$, since  $\int f ~ dg \wedge dh = \int g ~ dh \wedge df = \int h ~ df \wedge dg$. 
%

A non-canonical Hamiltonian formulation
is obtained by evaluating the functional derivative $E_{\zeta}$.
This yields a Poisson bracket
\begin{equation}
\{F,H\} = - \int E_{\zeta} J \left( H_{\zeta}, F_{\zeta} \right) dA  
\label{eq:PB}
\end{equation}
with the Casimir functional $E$.

\subsection{Physical interpretation of the conservation laws H and E}

The dynamics derived so far is not based on a physical interpretation of the two CLs $H$ and $E$. Consider a two-dimensional, nondivergent flow $(u,v)$ in the $(x,y)$ plane, derived from the stream-function $\psi$ that satisfies  
$ u = - \partial_y \psi, v = \partial_x \psi$.
%
%
With the vorticity
\begin{equation}
\zeta = \frac{\partial v}{\partial x} - \frac{\partial u}{\partial y} = \nabla^2 \psi ,
\label{eq:311}
\end{equation}
the first CL is the kinetic energy
\begin{equation}
H = \frac{1}{2} \int \left( u^2 + v^2 \right) dA .
\label{eq:314}
\end{equation}  
For periodic boundary conditions,
\begin{equation}
H = -  \frac{1}{2} \int \zeta \psi  dA ,
\label{eq:315}
\end{equation}  
and the functional derivative is
\begin{equation}
\frac{\delta H}{\delta \zeta} = - \psi .
\label{eq:316}
\end{equation}  
The kinetic energy $H$ play thus a unique role in hydrodynamics
since it is responsible for the advection in Eulerian flows
through the functional derivative with respect to $\zeta$. 

The second CL is the enstrophy 
\begin{equation}
E = \frac{1}{2} \int \zeta^2 dA ,
\label{eq:312}
\end{equation}
so that
\begin{equation}
\frac{\delta E}{\delta \zeta} = \zeta .
\label{eq:313}
\end{equation}

Using (\ref{eq:316}) and (\ref{eq:313}), (\ref{eq:36}) yields
\begin{equation}
\frac{\partial \zeta}{\partial t} = - J \left( \psi, \zeta \right) .
\label{eq:317}
\end{equation}
It should be noted that due to the Lagrangian conservation of  $\zeta$,  the system conserves arbitrary functionals $E_h = \int h( \zeta ) dA $. As noted by \cite{NevirBlender1993}, (\ref{eq:317}) can be rewritten in terms of any Casimir $E_h$, provided an opportune rescaling of the 2D Jacobian. 
%


The Nambu bracket (\ref{eq:38}), derived for the Euler equation, 
holds also for generalized Euler equations in which the stream-function $\psi$ is linked to an active tracer $\zeta$  by a fractional Poisson equation
 \cite{pierrehumbertetal94, Iwayama2010}
\begin{equation}
\zeta = \nabla^{\alpha} \psi .
\label{eq:322}
\end{equation}
In (\ref{eq:322}), the parameter $\alpha$  defines the degree of smoothing. 
While for $\alpha =2$, (\ref{eq:322}) implies the 2D Euler equation, for  $\alpha=1$  a physical model  is 
given by a stratified and rotating flow in a semi-infinite three-dimensional
domain with nonlinear advection of potential temperature at one of the boundaries and vanishing potential vorticity in the interior
\cite{blumen78,heldetal95}. The so obtained dynamics is called surface quasi-geostrophic (SQG) approximation. If the horizontal coordinates are a Legendre transform of the physical coordinates $(x,y)$ which follow the flow, and neglecting an arising nonlinear term, the surface semi-geostrophic approximation is obtained \cite{badin13}. 
Other cases physically realizable include the case with negative exponent $\alpha=-2$,
which corresponds to the limit of the large-scale quasi-geostrophic dynamics \cite{larichevmcwilliams91}, and the case $\alpha=3$ which represents a rotating shallow flow, which is the limit of a mantle convection model \cite{weinsteinetal89}.
Other values of $\alpha$ can be used for the study of systems characterized by different degrees of locality of the resulting turbulence \cite{smithetal02}. 
Noticeably, all the models defined by (\ref{eq:322}) have the same Nambu bracket (\ref{eq:38}) and CLs, with the only difference in the physical meaning of the CLs.

\section{Geometry of coupled continuous systems}
\label{sec:coupled}

  \subsection{Nambu bracket for a system with two degrees of freedom}

Consider now a system with two degrees of freedom $\zeta$ and $\mu$. Two generic CLs are considered, the first is the total energy, given by the sum of the kinetic energy and a potential energy which is  a functional of $\mu$ only
\begin{equation}
H = - \frac{1}{2} \int \zeta \psi dA + P[ \mu ] .
\label{eq:41}
\end{equation}				

The second CL is a bilinear coupling of vorticity to the 
second degree of freedom $\mu$ 
\begin{equation}
C = \int \zeta \mu dA .
\label{eq:42}
\end{equation}	

The time derivatives of the two CLs are
\begin{equation}
\frac{d H}{d t} = \int \left[ H_{\zeta} \frac{\partial \zeta}{\partial t} + H_{\mu} \frac{\partial \mu}{\partial t} \right] dA = 0 ,
\label{eq:43}
\end{equation}	
\begin{equation}
\frac{d C}{d t} = \int \left[ C_{\zeta} \frac{\partial \zeta}{\partial t} + C_{\mu} \frac{\partial \mu}{\partial t} \right] dA = 0 .
\label{eq:44}
\end{equation} 
The advection of vorticity $\zeta$  by the flow $\psi$ can be written as
\begin{equation}
\frac{\partial \zeta}{\partial t} ~ dx \wedge dy = d H_{\zeta} \wedge d C_{\mu} .
\label{eq:45}
\end{equation}
With this term the first part of the integral in  (\ref{eq:43}) vanishes.
%
%
The advection of $\mu$ is represented by 
\begin{equation}
\frac{\partial \mu}{\partial t} dx \wedge dy = - dC_{\zeta} \wedge dH_{\zeta} .
\label{eq:47}
\end{equation}
To satisfy $dH / dt = 0$, an additional term $Z$ is necessary
in $\partial \zeta / \partial t$, so that 
\begin{equation}
\frac{\partial \zeta}{\partial t} dx \wedge dy 
=  dH_{\zeta} \wedge dC_{\mu} + Z .
\label{eq:48}
\end{equation}
With  (\ref{eq:47}) and (\ref{eq:48}), and the identities
\begin{equation}
\begin{array}{ccc}
\int d \left( C_{\mu} C_{\zeta} \right) \wedge dH_{\zeta} =0 & and & \int d \left( H_{\mu} H_{\zeta} \right) \wedge dC_{\zeta} =0 ,
\end{array}
\label{eq:49}
\end{equation}
(\ref{eq:43}) becomes
\begin{equation}
\frac{dH}{dt} = \int H_{\zeta} \left( dH_{\zeta} \wedge dC_{\mu} + Z \right) dA + \int H_{\mu} dH_{\zeta} \wedge dC_{\zeta} dA = 0 ,
\label{eq:410}
\end{equation}
%
which requires that
\begin{equation}
Z = - dC_{\zeta} \wedge dH_{\mu} .
\label{eq:411}
\end{equation}
Using this $Z$-term, the derivatives $dH / dt$  and  $dC / dt$ vanish, and
we have finally 
\begin{equation}
\frac{\partial \zeta}{\partial t} dx \wedge dy = - dC_{\mu} \wedge dH_{\zeta} - dC_{\zeta} \wedge dH_{\mu} .
\label{eq:412}
\end{equation}
Notice that the additional $Z$ term does not contribute to $dC / dt$. This is a direct consequence of Nambu mechanics having the property of possessing hidden symmetries resulting in extra integrals of motion.
%
%
Using the forms obtained for $\partial \zeta / \partial t$  and  $\partial \mu / \partial t$, we can define a Nambu bracket for an arbitrary functional $F[\zeta, \mu]$ by including the corresponding functional derivatives $F_{\mu}$ and $F_{\zeta}$ 
\begin{equation}
\frac{d F}{d t} = \{ F, C, H \} 
\end{equation}
\begin{equation}
\{ F, C, H \} = - \int \left[ F_{\mu} dC_{\zeta} \wedge dH_{\zeta} + F_{\zeta} dC_{\zeta} \wedge dH_{\mu} + F_{\zeta} dC_{\mu} \wedge dH_{\zeta} \right]
\label{eq:414}
\end{equation}
As noted by \cite{Bihlo2008}, this Nambu bracket is the continuous analogue of the Nambu bracket for the heavy top, for which the fundamental Jacobi identity is known to be satisfied.

  \subsection{Example: Rayleigh-B\'{e}nard convection}

To demonstrate how the method can be applied in a practical context, 
the Rayleigh-B\'{e}nard convection is used as an example. The system has two dynamic variables, 
$\zeta$ and $\mu$, the first being the vorticity and 
the second represents a vertical temperature anomaly 
that interacts with the convective motion. 
The dynamic equations are \cite{Bihlo2008}
\begin{equation}
\frac{\partial \zeta}{\partial t} = J(\zeta , \psi) + \frac{\partial \mu}{\partial x} 
\label{eq:417}
\end{equation} 
\begin{equation}
\frac{\partial \mu}{\partial t} = J(\mu , \psi) + \frac{\partial \psi}{\partial x} ,
\label{eq:418}
\end{equation} 
where $\psi$ is the stream-function for the non-divergent flow $(u,v)$ in the vertically oriented $x$-$y$-plane, $u = - \partial \psi / \partial y, ~ v = \partial \psi / \partial x$. 
The two CLs are 
\begin{equation}
H = - \int \left( \frac{1}{2} \zeta \psi + \mu y \right) dx dy ,
\label{eq:419}
\end{equation} 
\begin{equation}
C = \int \zeta \left( \mu - y \right) dx dy ,
\label{eq:420}
\end{equation}   
where  $H$ is the total energy (kinetic and potential) and $C$ is based on Kelvin's circulation theorem. 

The dynamic equations for an arbitrary functional $F[\zeta, \mu]$ 
can be obtained by the Nambu bracket (\ref{eq:414}) 
(see also \cite{Bihlo2008,SalazarKurgansky2010}).
This example shows that the geometric approach based on 
orthogonality conditions (\ref{eq:43})-(\ref{eq:44}) yields a rather simple and 
straightforward method to construct the Nambu bracket. 

\section{Constitutive conservation laws}
\label{sec:ccl}

Following an idea already included in the original paper by Nambu \cite{Nambu1973}, 
constitutive conservation laws (CCLs) have been introduced in Nambu dynamics by \cite{NevirSommer2009}. The corresponding brackets have been classified by \cite{SalazarKurgansky2010} as Nambu brackets of type II. The main idea  
is the partitioning of a physical system into 
sub-systems, with conservation laws satisfied in the subsystem only. For the coupled system introduced in Section \ref{sec:coupled}, this means that $H$ and $C$ are conserved, but the dynamics is written in terms of two functionals which replace $C$ and are not conserved in the complete system. 
The resulting dynamics is split into two brackets: 
a first bracket, which uses enstrophy $E$ and energy $H$ 
and corresponds to 2D-hydrodynamics, without contributions from the buoyancy. 
The second bracket uses total energy $H$ and a conservation law $B$, 
which is based on the thermodynamics of the system only.
The system is decomposed by the dependencies  $E = E[\zeta]$  and  $B = B[\mu]$. It is convenient to use quadratic integrals 
\begin{equation}
E = \frac{1}{2} \int \zeta^2 dA , \qquad B = \frac{1}{2} \int \mu^2 dA .
\label{eq:51}
\end{equation} 
%
Notice that $B$ defined  here differs from the total buoyancy used by \cite{SalazarKurgansky2010}.
The overarching role of the total energy $H$ is 
expressed by $H = H[\zeta , \mu]$,  which combines kinetic and potential energy. 
It should be noted that the energy $H$ remains in all brackets. 

The conservation laws for $H$ and $C$ are the same as in (\ref{eq:43})-(\ref{eq:44}). To proceed with the determination of the Nambu bracket, the functional derivatives of $C$ are substituted by derivatives of $E$ and $B$
\begin{equation}
\begin{array}{cc}
C_{\zeta} = B_{\mu}, & C_{\mu} = E_{\zeta} ,
\end{array}
\label{eq:53}
\end{equation} 
thus the dynamical equations read as
\begin{equation}
\frac{\partial \zeta}{\partial t} dx \wedge dy 
= dH_{\zeta} \wedge dE_{\zeta} - dB_{\mu} \wedge dH_{\mu} ,
\label{eq:54}
\end{equation} 
\begin{equation}
\frac{\partial \mu}{\partial t} dx \wedge dy = dH_{\zeta} \wedge dB_{\mu} .
\label{eq:55}
\end{equation} 

Using (\ref{eq:54})-(\ref{eq:55}), one can formulate the tendency of an arbitrary functional $F[\zeta, \mu]$ 
\begin{equation}
\frac{d F}{d t} = - \int F_{\zeta} dE_{\zeta} \wedge dH_{\zeta} - \int F_{\zeta} dB_{\mu} \wedge dH_{\mu} - \int F_{\mu} dB_{\mu} \wedge dH_{\zeta} .
\label{eq:56}
\end{equation} 

The first term is the 2D bracket for incompressible hydrodynamics, which involves only $\zeta$-derivatives 
(\ref{eq:38})
\begin{equation}
\{ F,E,H \}_{\zeta \zeta \zeta} = - \int F_{\zeta} dE_{\zeta} \wedge dH_{\zeta} ,
\label{eq:57}
\end{equation} 
with the enstrophy $E$ and the energy $H$. The second bracket is
\begin{equation}
\{ F,B,H \}_{\zeta \mu \mu} = - \int F_{\zeta} ~ dB_{\mu} \wedge dH_{\mu} - \int F_{\mu} ~ dB_{\mu} \wedge dH_{\zeta} .
\label{eq:58}
\end{equation} 
The dynamics is split into two brackets with the CCLs $E$ and $B$, so that  
\begin{equation}
\frac{d F}{d t} =  \{ F,E,H \}_{\zeta \zeta \zeta} + \{ F,B,H \}_{\zeta \mu \mu} .
\label{eq:59}
\end{equation} 
In the classification introduced by \cite{SalazarKurgansky2010}, this is a Nambu bracket of type II. 
If buoyancy is neglected this decomposition yields 
a representation of the coupled system in terms of only 2D-hydrodynamics.  
%

\section{Summary}

In this study, Nambu systems with two CLs are reconsidered. For three degrees of freedom, the standard Nambu form is derived by a pure geometric approach in three-dimensional space, with the resulting dynamics given as a flow along the intersection of the two manifolds determined by the conserved quantities. This approach is quite general, since it uses only CLs and their dependencies on the dynamic variable and does not need the beforehand specification of the dynamic equations that are used. 

This geometric approach is transferred to a continuous system in 2D with two CLs. The derivation is based on two-forms with vanishing integrals in a periodic domain. 
Similar to the finite dimensional case, the resulting dynamics is constrained by an orthogonality condition. The result is a Nambu bracket which appears in two-dimensional hydrodynamics when  the dynamic variable is the vorticity and the functional derivatives of the two CLs are the vorticity and the streamfunction of the flow. This suggests an analogy in the geometry of the hydrodynamics equations and the equation for a Nambu triplet with finite degrees of freedom. 
The Nambu brackets holds its form also for the generalized Euler equation, which is defined by a fractional Poisson equation for the stream-function and an active scalar. 
Generalized Euler equations are interesting for the study of the formation of fronts and singularities \cite{constantinetal94}, 
conservative numerical algorithms based on the Nambu bracket \cite{Salmon2005} can thus be useful to  
avoid spurious accumulation or dissipation of energy and enstrophy.

Finally, a coupled system with two degrees of freedom is considered to demonstrate a further application of the method presented here. The model is analogous to Rayleigh-B\'{e}nard convection with vorticity and a temperature anomaly as dynamic variables. The approach is extended to construct a Nambu type II formulation using constitutive CLs. 
In the Rayleigh-B\'{e}nard example, the method yields two brackets: a first representing inviscid 2D hydrodynamics, and a second representing the coupling between hydrodynamics and thermodynamics. In the brackets the Hamiltonian is present in all terms and allows for a non-canonical Hamiltonian description of the system.

\section{Note}
Appeared as:\\
R. Blender and G. Badin, 2015: "Hydrodynamic Nambu mechanics derived by geometric constraints", Journal of Physics A: Mathematical and Theoretical, 48, 105501


\clearpage

\bibliographystyle{unsrt}

\begin{thebibliography}{10}

\bibitem{Nambu1973}
Y.~Nambu.
\newblock Generalized {H}amiltonian dynamics.
\newblock {\em Phys. Rev. D}, 7:2405--2412, 1973.

\bibitem{Takhtajan1994}
L.~Takhtajan.
\newblock On foundation of the generalized {N}ambu mechanics.
\newblock {\em Commun. Math. Phys.}, 160:295--315, 1994.

\bibitem{NevirBlender1994}
P.~N{\'e}vir and R.~Blender.
\newblock Hamiltonian and {N}ambu representation of the non-dissipative
  {L}orenz equations.
\newblock {\em Contrib. to Atmosph. Phys.}, 67:133--140, 1994.

\bibitem{Roupas2012}
Z.~Roupas.
\newblock Phase space geometry and chaotic attractors in dissipative {N}ambu
  mechanics.
\newblock {\em J. Phys. A: Math. Theor.}, 45:195101, 2012.

\bibitem{Chatterjee1996}
R.~Chatterjee.
\newblock Dynamical {S}ymmetries and {N}ambu {M}echanics.
\newblock {\em Letters Math. Phys.}, 36:117--126, 1996.

\bibitem{Cohen1975}
I.~Cohen.
\newblock Generalization of {N}ambu's {M}echanics.
\newblock {\em Int. J. Theor. Phys.}, 12:69--78, 1975.

\bibitem{NevirBlender1993}
P.~N\'{e}vir and R.~Blender.
\newblock A {N}ambu representation of incompressible hydrodynamics using
  helicity and enstrophy.
\newblock {\em J. Phys. A: Math. Gen.}, 26:1189--1193, 1993.

\bibitem{Salmon2008}
R.~Salmon.
\newblock {\em Lectures on {G}eophysical {F}luid {D}ynamics}.
\newblock Oxford University Press, Oxford, 2008.

\bibitem{Salmon2005}
R.~Salmon.
\newblock A general method for conserving quantities related to potential
  vorticity in numerical models.
\newblock {\em Nonlinearity}, 18:1--16, 2005.

\bibitem{Salmon2007}
R.~Salmon.
\newblock A general method for conserving energy and potential enstrophy in
  shallow water models.
\newblock {\em J. Atmos. Sci.}, 64:515--531, 2007.

\bibitem{SommerNevir2009}
M.~Sommer and P.~N\'{e}vir.
\newblock A conservative scheme for the shallow-water system on a staggered
  geodesic grid based on a {N}ambu representation.
\newblock {\em Quart. J. Roy. Meteorol. Soc.}, 135:485--494, 2009.

\bibitem{Bihlo2008}
A.~Bihlo.
\newblock Rayleigh-{B}\'{e}nard convection as a {N}ambu-metriplectic problem.
\newblock {\em J. Phys. A: Math. Theor.}, 41:292001, 2008.

\bibitem{GassmannHerzog2008}
A.~Gassmann and H.-J. Herzog.
\newblock Towards a consistent numerical compressible non-hydrostatic model
  using generalized {H}amiltonian tools.
\newblock {\em Quart. J. Roy. Meteorol. Soc.}, 134:1597--1613, 2008.

\bibitem{NevirSommer2009}
P.~N\'{e}vir and M.~Sommer.
\newblock Energy-vorticity theory of ideal fluid mechanics.
\newblock {\em J. Atmos. Sci.}, 66:2073--2084, 2009.

\bibitem{SalazarKurgansky2010}
R.~Salazar and M.~V. Kurgansky.
\newblock Nambu brackets in fluid mechanics and magnetohydrodynamics.
\newblock {\em J. Phys. A}, 43:305501, 2010.

\bibitem{BlenderLucarini2013}
R.~Blender and V.~Lucarini.
\newblock Nambu representation of an extended {L}orenz model with viscous
  heating.
\newblock {\em Physica D}, 243:86--91, 2013.

\bibitem{Sommeretal2011}
M.~Sommer, K.~Brazda, and M.~Hantel.
\newblock Algebraic construction of a {N}ambu bracket for the two-dimensional
  vorticity equation.
\newblock {\em Phys. Lett. A}, 375:3310--3313, 2011.

\bibitem{Fecko1992}
M.~Fecko.
\newblock On a geometrical formulation of the {N}ambu dynamics.
\newblock {\em J. Math. Phys.}, 33:926--929, 1992.

\bibitem{Arnold89}
V.I. Arnold.
\newblock {\em Mathematical {M}ethods of {C}lassical {M}echanics}.
\newblock Springer, New York, 1989.
\newblock 2nd ed.

\bibitem{pierrehumbertetal94}
R.T. Pierrehumbert, I.M. Held, and K.L. Swanson.
\newblock Spectra of local and nonlocal two-dimensional turbulence.
\newblock {\em Chaos Solit. Fract.}, 4:1111--1116, 1994.

\bibitem{Iwayama2010}
T.~Iwayama and T.~Watanabe.
\newblock Green's function for a generalized two-dimensional fluid.
\newblock {\em Phys. Rev. E}, 82:036307, 2010.

\bibitem{blumen78}
W.~Blumen.
\newblock Uniform potential vorticity flow: {P}art {I}. {T}heory of wave
  interactions and two-dimensional turbulence.
\newblock {\em J. Atmos. Sci.}, 35:774--783, 1978.

\bibitem{heldetal95}
I.~M. Held, R.~T. Pierrehumbert, S.~T. Garner, and K.~L. Swanson.
\newblock Surface quasi-geostrophic dynamics.
\newblock {\em J. Fluid Mech.}, 282:1--20, 1995.

\bibitem{badin13}
G.~Badin.
\newblock Surface semi-geostrophic dynamics in the ocean.
\newblock {\em Geophys. Astrophys. Fluid Dyn.}, 107:526--540, 2013.

\bibitem{larichevmcwilliams91}
V.D. Larichev and J.C. McWilliams.
\newblock Weakly decaying turbulences in an equivalent-barotropic fluid.
\newblock {\em Phys. Fluids}, 3:938--950, 1991.

\bibitem{weinsteinetal89}
S.A. Weinstein, P.L. Olson, and D.A. Yuen.
\newblock Time-dependent large aspect-ratio thermal convection in the {E}arth's
  mantle.
\newblock {\em Geophys. Astrophys. Fluid Dyn.}, 47:157--197, 1989.

\bibitem{smithetal02}
K.~S. Smith, G.~Boccaletti, C.C. Henning, I.~Marinov, C.Y. Tam, I.M. Held, and
  G.~K. Vallis.
\newblock Turbulent diffusion in the geostrophic inverse cascade.
\newblock {\em J. Fluid Mech.}, 469:13--48, 2002.

\bibitem{constantinetal94}
P.~Constantin, A.J. Majda, and E.~Tabak.
\newblock Formation of strong fronts in the {2-D} quasigeostrophic thermal
  active scalar.
\newblock {\em Nonlinearity}, 7:1495--1533, 1994.

\end{thebibliography}


\end{document}